\documentclass[
    ,final            
  ]
  {aipproc}

\layoutstyle{8x11double}

\begin{document}

\title{Pulsar Dreams}


\author{Jonathan Arons}{
  address={Astronomy Department, Physics Department and Theoretical Astrophysics Center \\
  University of California, Berkeley}
}

\begin{abstract}
 I share a few reminiscences and observations of 40 years of Pulsars.
\end{abstract}

\maketitle

The organizers of ``40 Years of Pulsars'' asked the Discussion Leaders to 
reflect on their own connection to pulsars' history in human affairs.

I was completely oblivious to pulsars' discovery, and not at all involved with their study at the time of their discovery.  In the winter of 1967-68,
I was obsessed with defining a thesis topic at Harvard, in what turned into
a study of why we can't see the intergalactic matter that was supposed to
provide the density needed to close - or at least flatten - the Universe.  That work included the first try at reionization of the Universe, by quasar UV rather than by the first stars. I don't
remember seeing the issue of Nature announcing pulsars' discovery, and I have no memory of conversations on the topic with other graduate students, or with faculty.  My earliest pulsar memory comes from attending an American Astronomical Society meeting in January, 1969 - characteristically, what
stuck in my mind was a piece of theoretical work, Peter Goldreich's invited talk in which he discussed his ideas on the force free magnetosphere. In that talk
he alluded to the possibility of pairs forming in the magnetosphere, something that has been one of my obsessions.  I didn't understand a word of what he
said.

I got to pulsars indirectly.  Having migrated to Princeton for a post-doc, 
I reacted with  Jerry Ostriker, forming a temporarily bound state in which we tried putting some substance into Jerry's belief that since he now understood pulsar spin down - vacuum strong electromagnetic dipole radiation - he wanted to understand quasars, which in that era were thought to be primarily emitters of 
non-thermal radiation - jazzed up Crab Nebulae.  Jerry's idea was that the cores of galaxies contained many neutron stars, all spinning down like the Crab, which would spit out the magnetic fields and high energy particles whose synchrotron emission would explain all that we see.  

Unfortunately, acceleration of test particles in strong vacuum waves does not give rise to an output particle spectrum whose synchrotron emission looks like any known source. The question was, would a collection of pulsars somehow behave differently than an isolated object.  Notice how we skipped over the obvious starting place, of understanding how
to account for the synchrotron nebulae around isolated objects in our own galaxy, before jumping into the quasar question - for me, an early lesson in theorists' chutzpah, go boldly even when one neither sees or understands the
path!  Being young and naive - I'm still naive, but certainly no longer young - I ignored the thought that we should start with what we could see in greater detail, and jumped into a multiple event model for quasars. We found some interesting physical effects that did suggest that particles in the overlapping 
strong waves could be accelerated into spectra more like what we thought were required by the observations of quasar SEDs.  By the time the resulting papers \cite{arons75, kulsrud75} got published in 1975, however,
the observers had come to realize that most quasars are more like thermal emitters (with spectra dominated by what is now known as the ``big blue bump''), and jets had been discovered to be a wide spread phenomenon, which were very hard to account for in a model powered by a hundred thousand separate stellar objects.

Freeman Dyson taught me another pulsar lesson while I was a postdoc.  In 1971 he recounted a wonderful tale, to the effect that more
than a decade earlier, he had been interested in white dwarfs' oscillations, to 
the point of collaborating with Bengt Str\"{o}mgren (at that time a member of the
Institute for Advanced Study's faculty), to do high speed optical photometry of 
Baade's star in the Crab Nebula, speculating that it might be an unusual
white dwarf.  However, being mentally focused on (obsessed with) white dwarfs,
they did their observations (I forget where) with a photometer with time resolution greater than one second, thus missing the 30 Hz optical pulsations
discovered more than a decade later by Cocke, Disney and Taylor \cite{cocke69}. Lesson taught, but still all too often unheeded - don't look only
where you expect to find something (``look under the lamppost''), quite possibly your basic assumptions are wrong; surprises wait, if only your search technique works outside the regime defined by the original question - a lesson pulsars have taught us again and again. 

But in the meantime, I became fascinated with the physical question of whether
pulsars do in fact spin down by magnetic dipole radiation. My starting point was
in asking what is a pulsar's mass loss rate, as a function of magnetic field and spin rate - in modern parlance, as a function of its voltage 
$\Phi = \sqrt{\dot{E}_R /c} \propto \sqrt{\dot{P}/P^3}$,
where $\dot{E}_R$ is the spin down luminosity. Ostriker and Gunn's answer
to the mass loss rate question was $\dot{M} = 0$; Goldreich and Julian's had
been $\dot{M} = m c \Phi /e$, where $m$ was either an electron mass or an ion mass depending on the sign of $\vec{\Omega} \cdot \vec{B}$ at the surface;
while Ruderman and Sutherland's, published in 1975, was a particle outflow several thousand times as dense as Goldreich's, in the form of electron-positron pairs.  

That pair creation model, which had a lot of interesting ideas as to how pulsars work in detail, sucked me in to pulsar research, and led away from the vacuum spindown model to an alternate theory of pair creation, and,
more importantly, to learning how to pay attention to the observations (sometimes) - ultimately leading to the realization that the young pulsar wind nebulae provide the essential observational data on pulsar $\dot{M}$, showing
that for systems with $\Phi > 10^{14}$ Volts, the plasma coming from these
neutron stars is dense - they are not vacuum rotators, or producers of
charge separated outflows - a force free model with quasi-neutral plasma seems to be the most useful starting place.  We still don't understand how they make as 
much plasma as they in fact do, and recent advances in constructing the force
free model have led to fundamental conflicts with the extant pair creation theory,
forcing serious rethinking of the modeling assumptions behind the last 30 years
of pulsar studies - see \cite{arons08}, which gives a discussion of this among a 
number of other issues in pulsar theory, a large, if still incomplete, 
list of references, as well as  an opinionated view of the immediate
prospects for pulsar research, from a theory perspective.

I hope for continued excitement, entertainment and fascination from pulsars 
for years to come, and am indebted to the organizers of ``40 Years of Pulsars''
for a most illuminating week spent in beautiful Montreal.

\begin{theacknowledgments}

My research efforts on pulsars and other topics has been supported by
NSF grant AST-0507813,  NASA grant NNG06G108G and DOE grant DE-FC02-06ER41453, all to the University of California, Berkeley; by the Department of Energy contract to the Stanford Linear Accelerator Center no. DE-AC3-76SF00515; and by the taxpayers of California.

\end{theacknowledgments}

\end{document}